\documentclass{svproc}
\usepackage{url}
\usepackage{amsmath}
\usepackage{amssymb}
\usepackage{graphicx}
\usepackage{epsfig}

\begin{document}
\mainmatter              
\title{Scale-invariant scalar field dark matter through the Higgs portal}
\titlerunning{Scale-invariant scalar field dark matter through the Higgs portal}  
%

\author{Catarina Cosme}

%
\authorrunning{Catarina Cosme} 
%
\tocauthor{Catarina Cosme}
\institute{Ottawa-Carleton Institute for Physics, Carleton University, 1125 Colonel By Drive, Ottawa, Ontario K1S 5B6, Canada\\
\email{ccosme@physics.carleton.ca}}


\maketitle              

\begin{abstract}

We introduce an oscillating scalar field coupled to the Higgs that can account for all dark matter in the Universe. Due to an underlying scale invariance of this model, the dark scalar only acquires mass after the electroweak phase transition. We discuss the dynamics of this dark matter candidate, showing that it behaves like dark radiation until the Electroweak phase transition and like non-relativistic matter afterwards. In the case of a negative coupling to the Higgs field, the scalar gets a vacuum expectation value after the electroweak phase transition and may decay into photons, although being sufficiently long-lived to account for dark matter. We show that, within this scenario, for a mass of 7 keV, the model can explain the observed galactic and extra-galactic 3.5 keV X-ray line. Nevertheless, it will be very difficult to probe this model in the laboratory in the near future. This proceedings paper is based on Refs. \cite{Cosme:2018nly,Cosme:2017cxk}.

\end{abstract}
\section{Introduction}
Dark matter (DM) is one of the greatest unsolved questions in Physics. This invisible form of matter constitutes almost 27\%
of the Universe’s energy density content and is required to explain its structure on large scales, the anisotropies in the Cosmic Microwave Background (CMB) and the galaxy rotation curves. Despite a large number of candidates that arise in theories beyond the Standard Model of Particle Physics (SM), the origin and the constitution of DM remain unknown. Although Weakly Interacting Massive Particles (WIMPs) are among the best-motivated thermally produced DM candidates, they have not been detected so far and the absence of new particles at the LHC motivates looking for alternatives to the WIMP paradigm.

In this work, we introduce an oscillating scalar field coupled to the Higgs as a dark matter candidate. Even though the Higgs-portal for dark matter has been explored in the context of thermal dark matter candidates (WIMPs), there are few proposals in the literature that investigate the case of a scalar field which is oscillating in the minimum of its quadratic potential, behaving like non-relativistic matter. Thus, we focus on a model where the oscillating scalar field dark matter obtains its mass only through the Higgs mechanism, i.e., through scale-invariant Higgs-portal interactions. We assume an underlying scale invariance of the theory, spontaneously broken by some mechanism that generates the Planck and the Electroweak scales in the Lagrangian, but which forbids a bare mass term for the dark scalar. The scale-invariance is maintained in the dark sector and, therefore, the dark scalar only gets mass after the Electroweak phase transition (EWPT).  Additionally, the model has a U(1) gauge symmetry which ensures the dark matter candidate stability if unbroken. The relevant interaction Lagrangian density is the following:
\begin{equation}
\mathcal{-L}_{int}=\pm\, g^{2}\left|\Phi\right|^{2}\left|\mathcal{H}\right|^{2}+\lambda_{\phi}\left|\Phi\right|^{4}+V\left(\mathcal{H}\right)+\xi R\left|\Phi\right|^{2}~,
\label{Lagrangian}
\end{equation}
where the Higgs potential, $V\left(\mathcal{H}\right)$, has the usual ``Mexican hat" shape, $g$ is the coupling between the Higgs and the dark scalar, $\lambda_{\phi}$ is the dark scalar's self-coupling and the last term in Eq. (\ref{Lagrangian}) corresponds to a non-minimal coupling of the dark matter field to curvature, where $R$ is the Ricci scalar and $\xi$ is a constant. 

In this paper, we will focus on the case where the Higgs-dark scalar interaction has a negative sign. Hence, the U(1) symmetry may be spontaneously broken, which can lead to interesting astrophysical signatures, as we will see later.

This proceedings paper is structured as follows: in section \ref{Dynamics before EWSB} , we describe the dynamics of the field from the inflationary period up to the EWPT. In section \ref{after EWSB} we discuss the behavior of the field after the EWPT, computing the present dark matter abundance. The phenomenology of this scenario is explored in section \ref{Pheno} and the conclusions are summarized in section \ref{Conc}. For more details and a complete list of references, see Refs. \cite{Cosme:2018nly,Cosme:2017cxk}.

\section{Dynamics before Electroweak symmetry breaking}
\label{Dynamics before EWSB}

In this section, we describe the evolution of the dark matter candidate before the EWPT, where the Higgs-portal coupling term has a negligible role. First, we discuss the dynamics of the dark scalar during the inflationary period, where the non-minimal coupling term dominates its behavior. Then, we examine the behavior of the field in the radiation era until the EWPT, where the self-interactions term drives the dark scalar dynamics.

\subsection{Inflation}
\label{inflation}

During inflation, in the regime where $\xi\gg g,\,\lambda_{\phi}$, the dynamics of the field is  mainly driven by the non-minimal coupling to gravity in Eq. (\ref{Lagrangian}). This term provides an effective mass to the dark scalar, $m_\phi$:
\begin{equation}
m_{\phi}\simeq\sqrt{12\,\xi}\,H_{inf}~,\label{mass inflation}
\end{equation}
where we have used the fact that the Ricci scalar during inflation is $R\simeq12\,H_{inf}^{2}$ and the Hubble parameter, written in terms of the tensor-to-scalar ratio $r$, reads:
\begin{equation}
H_{inf}\left(r\right)\simeq2.5\times10^{13}\left(\frac{r}{0.01}\right)^{1/2}\,\mathrm{GeV}~.\label{Hubble infl}
\end{equation}
Note that $m_{\phi}> H_{inf}$ for $\xi>1/12$. Thus, although the classical field is driven towards the origin during inflation, its average value never vanishes due to de Sitter quantum fluctuations on super-horizon scales. Any massive field during inflation exhibits quantum fluctuations that get streched and amplified by the Universe's expansion and, in particular, for  $m_{\phi}/H_{inf}>3/2$ ($\xi>3/16$) the amplitude of each Fourier mode with comoving momentum $k$ is given by \cite{Riotto:2002yw}:
\begin{equation}
\left|\delta\phi_{k}\right|^{2}\simeq\left(\frac{H_{inf}}{2\pi}\right)^{2}\left(\frac{H_{inf}}{m_{\phi}}\right)\frac{2\pi^{2}}{\left(a\,H_{inf}\right)^{3}}~,\label{Fourier modes m greater 3 over 2}
\end{equation}
where $a(t)$ is the scale factor. Integrating over the super-horizon comoving momentum $0<k<aH_{inf}$, at the end of inflation, the homogeneous field variance reads:
\begin{equation}
\left\langle \phi^{2}\right\rangle \simeq\frac{1}{3}\,\left(\frac{H_{inf}}{2\pi}\right)^{2}\frac{1}{\sqrt{12\xi}}~,\label{field variance very massive}
\end{equation}
which sets the initial amplitude for field oscillations in the post inflationary era:
\begin{equation}
\phi_{inf}=\sqrt{\left\langle \phi^{2}\right\rangle }\simeq\alpha\,H_{inf}\qquad\alpha\simeq0.05\,\xi^{-1/4}~.\label{initial amplitude after inflation very massive}
\end{equation}
After inflation, when $m_\phi \gg H$ is satisfied, the field oscillates about the minimum of its potential. Moreover, since  $R=0$ in a radiation-dominated era and $R\sim \mathcal{O}(H^2)$ in the following eras, we may neglect the effects of the non-minimal coupling term in the evolution of the field after inflation. Hence, we may conclude that the role of the non-minimal coupling to gravity is  to make the field sufficiently heavy during inflation so to suppress potential isocurvature modes in the CMB anisotropy spectrum.

\subsection{Radiation era}
\label{rad era}

After inflation and the reheating period (which we will assume to be instantaneous, for simplicity), the Universe becomes radiation-dominated and  $R=0$. Above the EWPT, the dominant term in the potential of the dark scalar is the quartic one (see Eq. (\ref{Lagrangian})), since the thermal effects can keep the Higgs field localized about its origin. The dark scalar acquires an effective field mass $m_{\phi}=\sqrt{3\,\lambda_{\phi}}\,\phi$ and, when the condition $m_\phi \gg H$ is satisfied, it starts to oscillate about the origin with an amplitude $\phi_{rad}$ given by:
\begin{equation}
\phi_{rad}  \left(T\right)=\frac{\phi_{inf}}{T_{rad}}\,T
 =\left(\frac{\pi^{2}\,g_{*}}{270}\right)^{1/4}\,\left(\frac{\phi_{inf}}{M_{Pl}}\right)^{1/2}\,\frac{T}{\lambda_{\phi}^{1/4}}~.\label{amplitude field radiation}
\end{equation}
where the temperature at the onset of fields oscillations, $T_{rad}$, reads
\begin{equation}
T_{rad}=\lambda_{\phi}^{1/4}\,\sqrt{\phi_{inf}\,M_{Pl}}\,\left(\frac{270}{\pi^{2}\,g_{*}}\right)^{1/4}~,\label{T rad}
\end{equation}
$g_{*}$ is the number of relativistic degrees of freedom and $M_{Pl}$ is the reduced Planck mass. Since the dark scalar's amplitude decays as $a^{-1}\propto T$ and $\rho_{\phi}\sim a^{-4}$, we conclude that the field behaves like dark radiation during this period.

As soon as the temperature of the Universe drops below the Electroweak scale, both the Higgs and the dark scalar fields acquire a vacuum expectation value (vev) and, consequentely, the Higgs generate a mass for the dark scalar, as we will see in the next section. The Electroweak phase transition is completed when the leading thermal contributions to the Higgs potential become Boltzmann-suppressed,  at approximately  $T_{EW}\sim m_{W}$, where $m_{W}$ is the $W$ boson mass.

\section{Dynamics after the Electroweak symmetry breaking}
\label{after EWSB}

At the EWPT, the relevant interaction potential is:
\begin{equation}
V(\phi,h)=-\,\frac{g^{2}}{4}\,\phi^{2}\,h^{2}+\frac{\lambda_{\phi}}{4}\,\phi^{4}+\frac{\lambda_{h}}{4}\,\left(h^{2}-\tilde{v}^{2}\right)^{2},~\label{Lagrangian neg}
\end{equation}
where the Higgs self-coupling is $\lambda_{h}\simeq0.13$ .

At this point, the Higgs and the dark scalar acquire a non-vanishing vev, respectively:
\begin{equation}
h_{0}=\left(1-\frac{g^{4}}{4\,\lambda_{\phi}\,\lambda_{h}}\right)^{-1/2}\tilde{v}\equiv\mathrm{v},\qquad\phi_{0}=\frac{g\,\mathrm{v}}{\sqrt{2\lambda_{\phi}}}~,\label{vevs}
\end{equation}
where $\mathrm{v}=246\,\mathrm{GeV}$. Notice that a non-vanishing vev for the dark scalar implies $g^4<4\lambda_\phi \lambda_h$, which we assume to hold. The mass of the dark scalar, which is generated only by the Higgs, is then:
\begin{equation}
m_{\phi}=g\,\mathrm{v}~.\label{mass neg gv}
\end{equation}

As pointed out in Refs. \cite{Cosme:2018nly,Cosme:2017cxk}, the dark scalar starts to oscillate about $\phi_{0}$,  with an amplitude  $\phi_{DM}\equiv x_{DM} \,\phi_0$ with $x_{DM}\lesssim1$ once the leading contributions to the Higgs potential become Boltzmann suppressed, below $T_{EW}\sim m_{W}$. This  $x_{DM}$ is not an extra parameter of the model, it is just a theoretical uncertainty that takes into account the evolution of the dark scalar during the Electroweak crossover. Although a numerical simulation of the dynamics of the field during the Electroweak crossover would be required, we can estimate the value of $x_{DM}$. Since $T_{EW}\lesssim T_{CO}$ by an $\mathcal{O}\left(1\right)$ factor, where $T_{CO}$ corresponds to the Electroweak crossover temperature, and given that $\phi\sim T$ while behaving as radiation and $\phi\sim T^{3/2}$ while behaving as non-relativistic matter, the field's amplitude might decrease by at most an $\mathcal{O}\left(1\right)$ factor as well. For more details, see Refs. \cite{Cosme:2018nly,Cosme:2017cxk}. Hence, we may conclude that the field smoothly changes from dark radiation to a cold dark matter behavior at the EWPT, as its potential becomes quadratic about the minimum.

As soon as the dark scalar starts to behave like cold dark matter, its amplitude evolves with the temperature as $\phi\left(T\right)=\phi_{DM} (T/T_{EW})^{3/2}$ and the number of particles in a comoving volume, $\frac{n_{\phi}}{s}$, becomes constant: 
\begin{equation}
\frac{n_{\phi}}{s}=\frac{45}{4\pi^{2}g_{*S}}\frac{m_\phi\phi_{DM}^{2}}{T_{EW}^{3}}~,\label{n over s}
\end{equation}
where $g_{*S}\simeq 86.25$ is the number of relativistic degrees of freedom contributing to the entropy at $T_{EW}$,  $s=\frac{2\pi^{2}}{45}\,g_{*S}\,T^{3}$ is the entropy density of radiation and $n_{\phi}\equiv\frac{\rho_{\phi}}{m_{\phi}}$ is the dark matter number density. We can use this to compute the present DM abundance, $\Omega_{\phi,0}\simeq 0.26$, obtaining the following relation for the field's mass:
\begin{equation}
m_{\phi}=\left(6\,\Omega_{\phi,0}\right)^{1/2}\left(\frac{g_{*S}}{g_{*S0}}\right)^{1/2}\left(\frac{T_{EW}}{T_{0}}\right)^{3/2}\frac{H_{0}M_{Pl}}{\phi_{DM}},\label{mass neg}
\end{equation}
where $g_{*S0}$, $T_{0}$ and $H_{0}$ are the present values of the number of relativistic degrees of freedom, CMB temperature and
Hubble parameter, respectively. Then, plugging Eq. (\ref{mass neg gv}) into Eq. (\ref{mass neg}), we find a relation between $g$ and $\lambda_\phi$:
\begin{equation}
g\simeq2\times10^{-3}\,\left(\frac{x_{DM}}{0.5}\right)^{-1/2}\,\lambda_{\phi}^{1/4}~.\label{g lambda neg}
\end{equation}
This relation is a key point of our model: essentially, it has only a single free parameter, which we take to be the mass of the field. We will come back to this when discussing the phenomenology of the model. 

The idea of this work is to introduce a dark matter candidate which is never in thermal equilibrium with the cosmic plasma. However, there are two main processes that can lead to the evaporation of the condensate. One of them is the Higgs annihilation into higher-momentum $\phi$ particles, which is prevented if \cite{Cosme:2018nly,Cosme:2017cxk} 
\begin{equation}
g\lesssim8\times10^{-4}\,\left(\frac{g_{*}}{100}\right)^{1/8}~.\label{upper bound on g}
\end{equation}
The other process is the production of $\phi$ particles from the coherent oscillations of the background condensate in a quartic potential, which is not efficient if \cite{Cosme:2018nly,Cosme:2017cxk}
\begin{equation}
\lambda_{\phi}<6\times10^{-10}\left(\frac{g_{*}}{100}\right)^{1/5}\left(\frac{r}{0.01}\right)^{-1/5}\xi^{1/10}~.\label{upper bound on lambda}
\end{equation}
If the constraints of Eqs. (\ref{upper bound on g}) and (\ref{upper bound on lambda}) are satisfied, the dark scalar is never in thermal equilibrium with the cosmic plasma, behaving like an oscillating condensate of zero-momentum particles throughout its cosmic history. Eq.~(\ref{upper bound on lambda}) yields the most stringent constraint on the model, limiting the viable dark matter mass to be $m_\phi\lesssim1\,\mathrm{MeV}$ \cite{Cosme:2018nly,Cosme:2017cxk}. 

\section{Phenomenology}
\label{Pheno}

In this section, we will discuss two possible ways of probing the proposed model. For more examples and a complete and detailed discussion, see Refs. \cite{Cosme:2018nly,Cosme:2017cxk}.

\subsection{Dark matter decay}

Since the dark scalar and the Higgs field are coupled, they exhibit a small mass mixing, $\epsilon=\frac{g^{2}\,\phi_{0}\,\mathrm{v}}{m_{h}^{2}}~$ \cite{Cosme:2018nly}. This means that the dark scalar can decay into the same decay channels as the Higgs, provided that they are kinematically accessible. Due to the mass restriction coming from Eq.~(\ref{upper bound on lambda}), which translates into $m_\phi\lesssim1\,\mathrm{MeV}$, the only kinematically accessible decay channel is the decay into photons. It is possible to show that the decay witdth of the dark matter candidate into photons is suppressed by a factor $\epsilon^2$ with respect to the decay width of a virtual Higgs boson into photons, yielding for the dark scalar's lifetime \cite{Cosme:2018nly,Cosme:2017cxk}:
\begin{equation} \label{decay time}
\tau_{\phi}\simeq7\times10^{27}\left(\frac{7\,\mathrm{keV}}{m_{\phi}}\right)^{5}\,\left(\frac{x_{DM}}{0.5}\right)^{2}\,\mathrm{sec}.
\end{equation}
Hence, although the lifetime is much larger than the age of the Universe, it can lead to an observable monochromatic line in the spectrum of galaxies and galaxy clusters. 

\begin{figure}[htbp]
\begin{centering}
\includegraphics[scale=0.4]{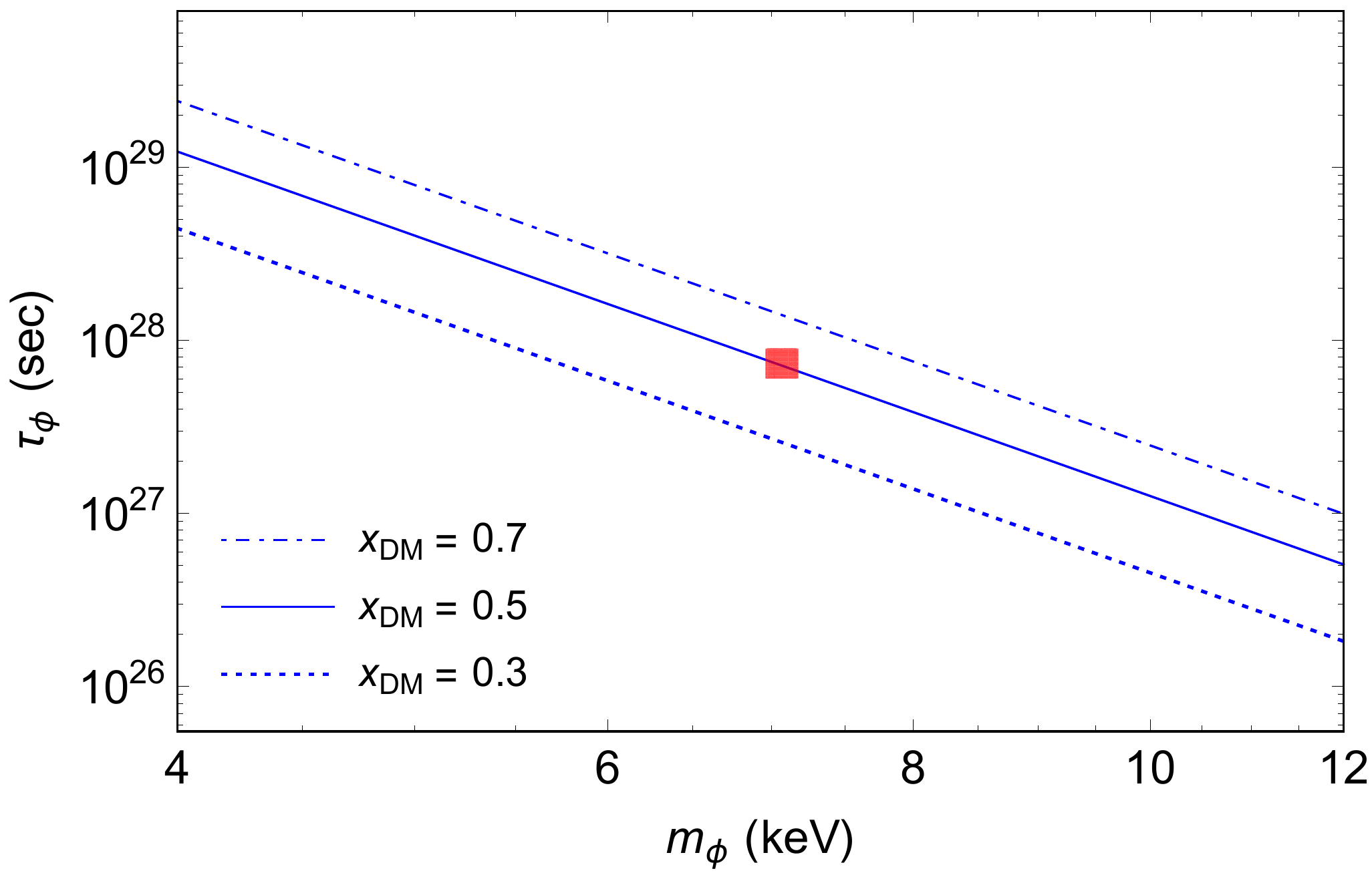}
\par\end{centering}
\caption{Lifetime of the scalar field dark matter as a function of its mass, for different values of $x_{DM}\lesssim 1$. The horizontal red band corresponds to the values of $\tau_\phi$ that can account for the 3.5 keV X-ray line detected by XMM-Newton for a mass around $7\,\,\mathrm{keV}$. From \cite{Cosme:2017cxk}. \label{plot}}
\end{figure}

Recently, the XMM-Newton X-ray observatory detected a line at 3.5 keV in the Galactic Center, Andromeda and Perseus cluster  \cite{Bulbul:2014sua,Boyarsky:2014jta,Boyarsky:2014ska,Cappelluti:2017ywp}. The nature of this line has arisen some interest in the scientific community, leading to several interesting proposals in the literature, in particular, the possibility of it resulting from DM decay or annihilation \cite{Cappelluti:2017ywp,Higaki:2014zua,Jaeckel:2014qea,Dudas:2014ixa,Queiroz:2014yna,Heeck:2017xbu}. In fact,  the analysis in Refs. \cite{Boyarsky:2014ska,Ruchayskiy:2015onc} has shown that the intensity of the line observed in the astrophysical systems mentioned above could be explained by the decay of a DM particle with a mass of $\simeq$ 7 keV and a lifetime in the range $\tau_{\phi}\sim\left(6-9\right)\times10^{27}$ sec. In the case of our dark scalar field model, fixing the field mass to this value, we predict a DM lifetime exactly in this range, up to some uncertainty in the value of the field amplitude after the EWPT parametrized by $x_{DM}\lesssim 1$. This is illustrated in Fig.~\ref{plot}. Notice that, for this mass,  $g\simeq3\times10^{-8}$ and $\lambda_{\phi}\simeq4\times10^{-20}$, satisfying the constraints in Eqs. (\ref{upper bound on g}) and (\ref{upper bound on lambda}).

The uniqueness of this result should be emphasized: our model predicts that the decay of the dark scalar $\phi$ into photons produces a 3.5 keV line compatible with the observational data, with effectively only one free parameter: either $g$ or $\lambda_{\phi}$. Recall that, originally, the model involves four parameters - the couplings $g$ and $\lambda_{\phi}$, the non-minimal coupling $\xi$ and the scale of inflation $r$. The role of $\xi$  is simply to suppress the potential cold dark matter isocurvature perturbations, while $r$ only sets the initial amplitude of the field at the beginning of the radiation era. At the EWPT, the field starts to oscillate around $\phi_{0}$, with an initial amplitude of this order - which does not depend on $\xi$ nor $r$. So, when the dark scalar starts to behave effectively as cold dark matter, only  $g$ and $\lambda_{\phi}$ affect its dynamics. Therefore, we have three observables that rely on just two parameters ($g$ and $\lambda_{\phi}$) -  the present dark matter abundance, the dark scalar's mass and its lifetime. Fixing the present dark matter abundance, we get a relation between $g$ and $\lambda_{\phi}$ (Eq.~(\ref{g lambda neg})), implying that $m_{\phi}$ and $\tau_{\phi}$ depend exclusively on the Higgs-portal coupling. Hence, the prediction for the magnitude of the 3.5 keV line in different astrophysical objects is quite remarkable and, as far as we are aware, it has not been achieved by other scenarios, where the dark matter's mass and lifetime can be tuned by different free parameters.


\subsection{Invisible Higgs decays into dark scalars}

One way to probe the Higgs-portal scalar field dark matter is to look for invisible Higgs decays into dark scalar pairs.  The corresponding decay width is:
\begin{equation}
\Gamma_{h\rightarrow\phi\phi}=\frac{1}{8\pi}\,\frac{g^{4}\mathrm{v^{2}}}{4\,m_{h}}\,\sqrt{1-\frac{4m_{\phi}^{2}}{m_{h}^{2}}}~,\label{decay H inv}
\end{equation}
where $m_{h}$ is the Higgs mass. Assuming the upper limit for the dark matter mass, $m_{\phi}=1\,\mathrm{MeV}$, the bound on the branching ratio is
\begin{equation}
Br\left(\Gamma_{h\rightarrow inv}\right)<10^{-19}~.\label{BR MeV}
\end{equation}
Considering that the current experimental limit is 
\begin{equation}
Br\left(\Gamma_{h\rightarrow inv}\right)=\frac{\Gamma_{h\rightarrow inv}}{\Gamma_{h}+\Gamma_{h\rightarrow inv}}\lesssim0.23~,\label{branching ratio}
\end{equation}
where we assume that $\Gamma_{h\rightarrow inv}=\Gamma_{h\rightarrow\phi\phi}$, we conclude that this process is too small to be measured with current technology. However, it may serve as motivation for extremely precise measurements of the Higgs boson's width in future collider experiments, given any other experimental or observational hints for light Higgs-portal scalar field dark matter, such as, for instance, the 3.5 keV line that we have discussed earlier.

\section{Conclusions}
\label{Conc}

In this proceedings paper, we summarize the results of  Refs. \cite{Cosme:2018nly,Cosme:2017cxk}, where we have shown that an oscillating scalar field coupled to the Higgs boson is a viable DM candidate that can explain the observed 3.5 keV X-ray line. This is a simple model, based on the assumed scale-invariance of DM interactions, and, at the same time, extremely predictive, with effectively only a single free parameter upon fixing the present DM abundance. Hence, our scenario predicts a 3.5 keV X-ray line with the observed properties for the corresponding value of the DM mass, although it will be very difficult to probe it in the laboratory in the near future.

%
%

\end{document}